\documentclass[12pt,preprint]{aastex}

\shorttitle{Night Sky Spectrum of Xinglong Observatory}
\shortauthors{Zhang et al.}

\begin{document}

\title{The Night Sky Spectrum of Xinglong Observatory: Changes from 2004 to 2015}

\author{JI-CHENG ZHANG$^{1,2}$, ZHOU FAN$^{2}$, JING-ZHI YAN$^{3,4}$, Yerra Bharat Kumar$^{2}$, HONG-BIN LI$^{2}$, DONG-YANG GAO$^{1}$, XIAO-JUN JIANG$^{2}$}
\affil {1.Shandong Provincial Key Laboratory of Optical Astronomy and Solar-Terrestrial Environment, Institute of Space Sciences, School of Space Science and Physics,Shandong University, Weihai, 264209, China, jczhang@bao.ac.cn}
\affil {2.Key Laboratory of Optical Astronomy, National Astronomical Observatories, Chinese Academy of Sciences, Beijing 100012, China}
\affil {3.Purple Mountain Observatory, Chinese Academy of Sciences, Nanjing, China}
\affil {4.Key Laboratory of Dark Matter and Space Astronomy, Chinese Academy of Sciences, Nanjing, China}

\begin{abstract}
  Spectroscopic measurements on the night sky of Xinglong Observatory for a period of 12 years from 2004 to 2015 are presented. The spectra were obtained in moonless clear nights by using OMR spectrograph mounted on 2.16-m reflector with wavelength coverage of 4000-7000 {\AA}. The night sky spectrum shows the presence of emission lines from Hg~{\sc i} and Na~{\sc i} due to local artificial sources, along with the atmospheric emission lines, i.e. O~{\sc i} and OH molecules, indicates the existence of light pollution. We have monitored the night sky brightness during whole night and found little decrement in the sky brightness with time, but the change is not significant. Also, we monitored the light pollution level in different azimuthal direction and found that the influence of light pollution from the direction of Beijing city is stronger compared with that from the direction of Tangshan city and others. The analysis from night sky spectra for the entire data set suggested that the zenith sky brightness of Xinglong Observatory has brightened about 0.5 mag arcsec$^{-2}$ in V and B bands from 2004 to 2015. We recommend consecutive spectroscopic measurements of the night sky brightness at Xinglong Observatory in future, not only for monitoring but also for scientific reference.
\end{abstract}


\section{INTRODUCTION}

Xinglong Observatory of National Astronomical Observatories, Chinese Academy of Sciences (NAOC) is one of the major optical observatories in China, located at a distance of 120~km towards North-East to Beijing, the capital city of China. The observatory hosts nine optical telescopes with apertures ranging from 0.5 to 4-m diameter. There are about 63$\%$ spectroscopic nights per year to perform observations in this site (See \citealt{2015PASP..127.1292Z}).

Night sky brightness is one of the fundamental parameters of an optical observatory that restricts the limiting magnitude for any planned observations. The brightness of moonless night sky are generated from natural sources mainly contributed by airglow, zodiacal light, and integrated starlight \citep{1998A&AS..127....1L}, and from artificial sources due to the lighting system of neighboring towns \citep{1999A&AS..140..345D}. With the economic development and population growth of the surrounding cities, their attendant light pollution also grows. \cite{1999AcApS..19..220J} presented an identification of the night sky emission lines of Xinglong Observatory with spectral coverage from 5300 to 8200 {\AA} during 1996-1998, and found that Na~{\sc i} and Hg~{\sc i} lines from artificial sources are quite weak.

Previous studies on sky brightness are mainly measured using broadband photometry. However, such measurements sometimes may be misleading, as it encompasses both natural airglow and artificial sources \citep{2000PASP..112..566M}. In order to better understand the contribution of atmospheric and artificial light sources, spectrophotometric measurements on sky brightness is suggested, which can distinguish the artificial sources from the natural sources clearly. \cite{2010PASP..122.1246N} presented a way to identify the contribution from specific elements that influence overall sky brightness.

The spectrophotometric measurements has been widely used at various international optical observatories. \cite{1999A&AS..140..345D} presented a survey in Venezuela and Italy using a small spectrograph with spectral coverage from 4100 to 6400 {\AA}. Night sky spectra of the Kitt Peak, during 1998, were analyzed with wavelength coverage from 3800 to 6500 {\AA} by \cite{1990PASP..102.1046M}. Then \cite{2000PASP..112..566M} presented an absolute spectrophotometry of the night sky from $\sim$3700 to 6700 {\AA} over two astronomical sites in southern Arizona, Kitt Peak and Mount Hopkins, and measured for different azimuthal directions and different zenith distances, then converted to broadband magnitudes and gave the comparison with the night sky spectra in 1988. \cite{2010PASP..122.1246N} presented new absolute spectrophotometry of the Kitt Peak night sky during 2009-2010, and they strove to use the same observation and data reduction manner of \citet{2000PASP..112..566M}, and compared with published data. \cite{2004JKAS...37...87S} presented the spectrophotometry of night sky over Bohyunsan Optical Astronomy Observatory (BOAO), which is located on top of Mount Bohyun, with the nearly entire visible wavelength from 3600 to 8600 {\AA}, and the authors compared the night sky spectrum with Kitt peak. Site testing for observatories also used the night sky spectra to analyze local light pollution (e.g. \citealt{2007PASP..119.1186S}, \citealt{2010PASP..122..363M}). These night sky studies are mainly based on relative low-resolution spectra, there may be different spectral coverage and blended lines. In order to identify as many lines as possible from the contribution of light pollution, \cite{2003PASP..115..869S} presented the night sky spectrum of light pollution at the Lick Observatory from 3800 to 9200 {\AA} with a high spectral resolution(R$\sim$45000), and identified a large variety of lines from light pollution.

In this work, we study the night sky brightness for a period of 12 years through the spectroscopic measurements in visible wavelength. This paper consists of four sections. Section~1 gives a brief introduction of Xinglong Observatory and the research basis of night sky spectra around the world. Section~2 describes the details of data acquisition and reduction. Analysis of the night sky spectrum at Xinglong Observatory and its results are presented in Section~3. We discuss our results and conclude in Section~4.

\section{DATA ACQUISITION AND REDUCTION}

The spectral data used in our study was from the observations of 2.16-m reflector at Xinglong Observatory. The telescope equipped with three instruments; 1. Beijing Faint Object Spectrograph and Camera (BFOSC) available for imaging and low resolution spectroscopy; 2. Spectrograph made by Optomechanics Research Inc.(OMR) for low resolution spectroscopy; 3. Fiber-fed High Resolution Spectrograph (HRS). Detailed introduction and up to date status of these three instruments are described in Fan et al. (2016) (in preparation). We searched the raw data from OMR and BFOSC archive for our analysis. As our study is byproduct of research interest of various observers who obtained data with different instruments, we need to select the appropriate data depending on our requirement through the logfiles of 2.16-m telescope recorded by astronomers in every observing night. This selection procedure is an essential and complicated work for this study. The main criteria for selecting the low resolution spectroscopic data are as follows: 1. Moonless clear nights with good astronomical seeing; 2. Exposure time should be at least 1800 seconds and 3600 seconds is ideal; 3. Preferable location of objects is at least $15^{\circ}$ away from the galactic plane as suggested by \cite{2000PASP..112..566M}. We ensure all the selected data were observed with same set up of spectrograph, leading to an accurate comparison. Finally, we decide to select the OMR data for our night sky spectrum analysis by considering the resolution, wide wavelength coverage and long time span.

OMR spectra are taken on a Princeton Instruments (PI) PIXIS 1340$\times$400 CCD with pixel size of 20 $\mu$m, and pixel scale of 0.96$''$. We selected the spectral data taken with 300 l/mm grating, which provides dispersion of 4.0{\AA}/pixel that covers required wavelength coverage of 4000-7000 {\AA}. As the mean and median seeing value of Xinglong Observatory over an year are 1.9$''$ and 1.7$''$, respectively (see \citealt{2015PASP..127.1292Z}), we selected slit width around 2.3$''$ in order to include more photons and improve the Signal to Noise Ratio (SNR). There are different lamps for OMR wavelength calibration and flatfield correction, here we used He-Ar lamp for our spectral wavelength calibration and halogen tungsten lamp for flatfield correction. Every night more than two spectroscopic standard stars were observed for flux calibration.

\begin{figure*}[htbp]
\centering
\includegraphics[angle=0,scale=0.35]{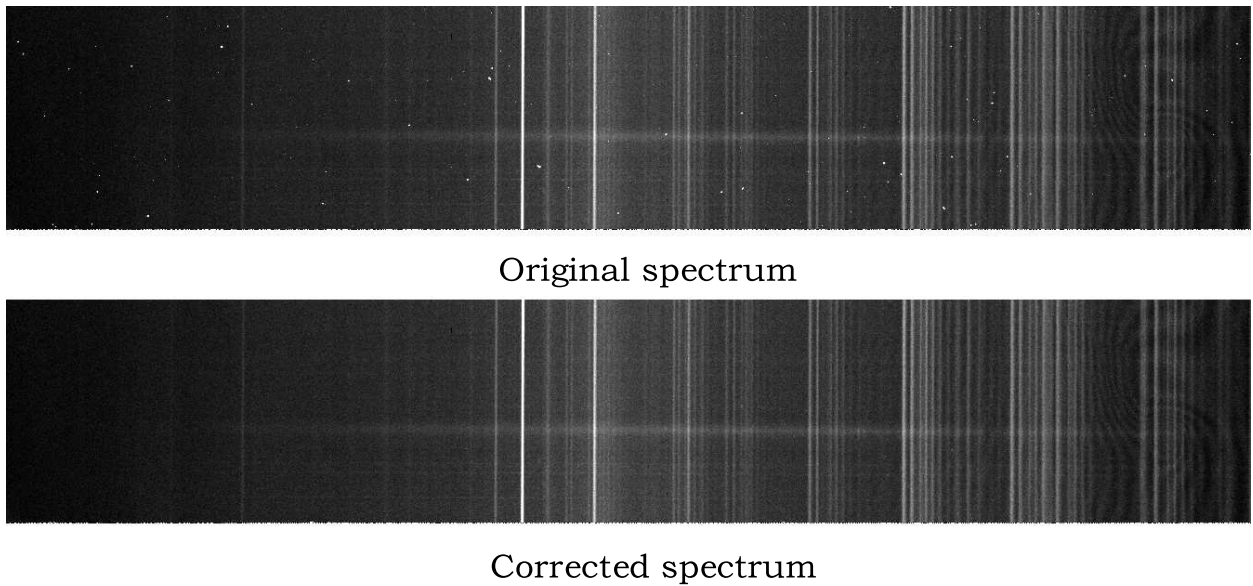}
\hfill \caption{Comparison of sample 2D spectrum before (upper panel) and after (lower panel) cosmic ray removal.}
\label{fit1}
\end{figure*}

Raw data were processed in the standard procedure using various available tasks in Image Reduction and Analysis Facility (IRAF) \footnote[1]{IRAF is distributed by National Optical Astronomy Observatory, which is operated by the Association of Universities for Research in Astronomy, Inc.(AURA) under cooperative agreement with the National Science Foundation.} and Interactive Data Language (IDL). As the dark current of CCD is negligible, we corrected for bias and flatfield to object images. The raw frames are contaminated with lot of cosmic rays due to long exposure times. To remove the influence of cosmic rays on the spectra of objects and night sky, we used the cosmic-ray rejection by Laplacian edge detection. Details on algorithms and introductions are presented in \cite{2001PASP..113.1420V}. Figure~\ref{fit1} shows the image before and after cosmic-ray removal. Upper panel is the original spectrum and lower panel is the corrected spectrum with no cosmic rays, which suggests Laplacian edge detection works well in removing the cosmic rays.

We have extracted the portion of night sky spectrum in the same frame and same dispersion direction of stellar spectrum. We have selected the "night sky" part as far as possible from stellar contribution in order to avoid the contamination from stellar lines. We also looked carefully the contribution from faint stars in the frames and removed using the standard tasks available in IRAF. As spectra of the object in CCD has slight dispersion curvature, we have performed the curvature correction to avoid errors in wavelength calibration. We have done flux calibration of observed spectroscopic standard stars and used it as reference to calibrate the night sky spectrum. The final spectra were corrected for extinction using local atmospheric extinction file. We converted the flux units to erg cm$^{-2}$s$^{-1}${\AA}$^{-1}$ for further analysis. As we are also interested in broadband of sky brightness, we have measured the magnitudes in broadband (Johnson system) B and V by convolving the night sky spectra with corresponding sensitivity curves \footnote[2]{http://www.aip.de/en/research/facilities/stella/instruments/data/johnson-ubvri-filter-curves.}.

\section{NIGHT SKY SPECTRUM AND ITS RESULTS}

\begin{figure*}[!htb]
\centering
\includegraphics[angle=0,scale=0.45]{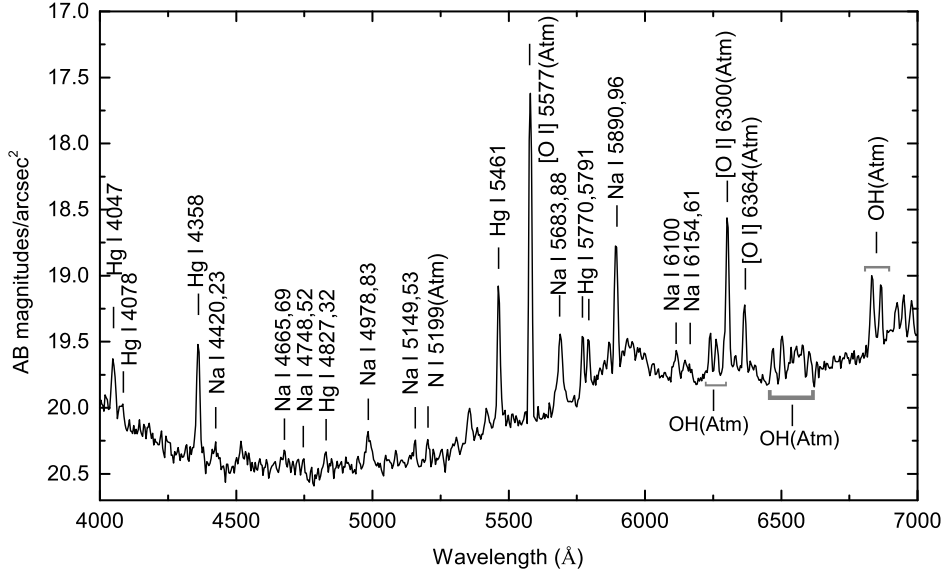}
\hfill \caption{A typical spectrum in the range of 4000 to 7000 {\AA} of night sky at Xinglong Observatory. Emission lines from natural (atmosphere) and artificial (Hg and Na) sources  are identified and marked.}
\label{fit2}
\end{figure*}

 The typical night sky spectrum of Xinglong Observatory is shown in Figure~\ref{fit2}, and the emission lines from artificial and natural light sources are identified and marked. The artificial sources are known to be Mercury (Hg) vapour lamp, low pressure Sodium (Na) vapour lamp (LPS), and high pressure Sodium (Na) vapour lamp (HPS). Prominent Hg~{\sc i} lines are noticed at 4047, 4358, 5461, 5770 and 5791 {\AA}, and weak lines of Hg~{\sc i} are noticed at 4078, 4827,4832 {\AA}. Na~{\sc i} emission lines at 4420, 4423, 4665, 4669, 4748, 4752, 4978, 4983, 5149, 5153, 6100, 6154 and 6161 {\AA} are weaker compared to stronger lines at 5683, 5688, 5890 and 5896 {\AA}. The strong Na emission lines in the region of 5500 - 5900 {\AA} have contribution from both LPS and HPS whereas other Na lines are from HPS. Oxygen emission lines in our spectra are mainly concentrated at 5577, 6300 and 6364 {\AA}. OH molecule lines are mainly distributed in the wavelength of redder than 6500 {\AA}.

\begin{figure*}[htbp]
\centering
\includegraphics[angle=0,scale=0.45]{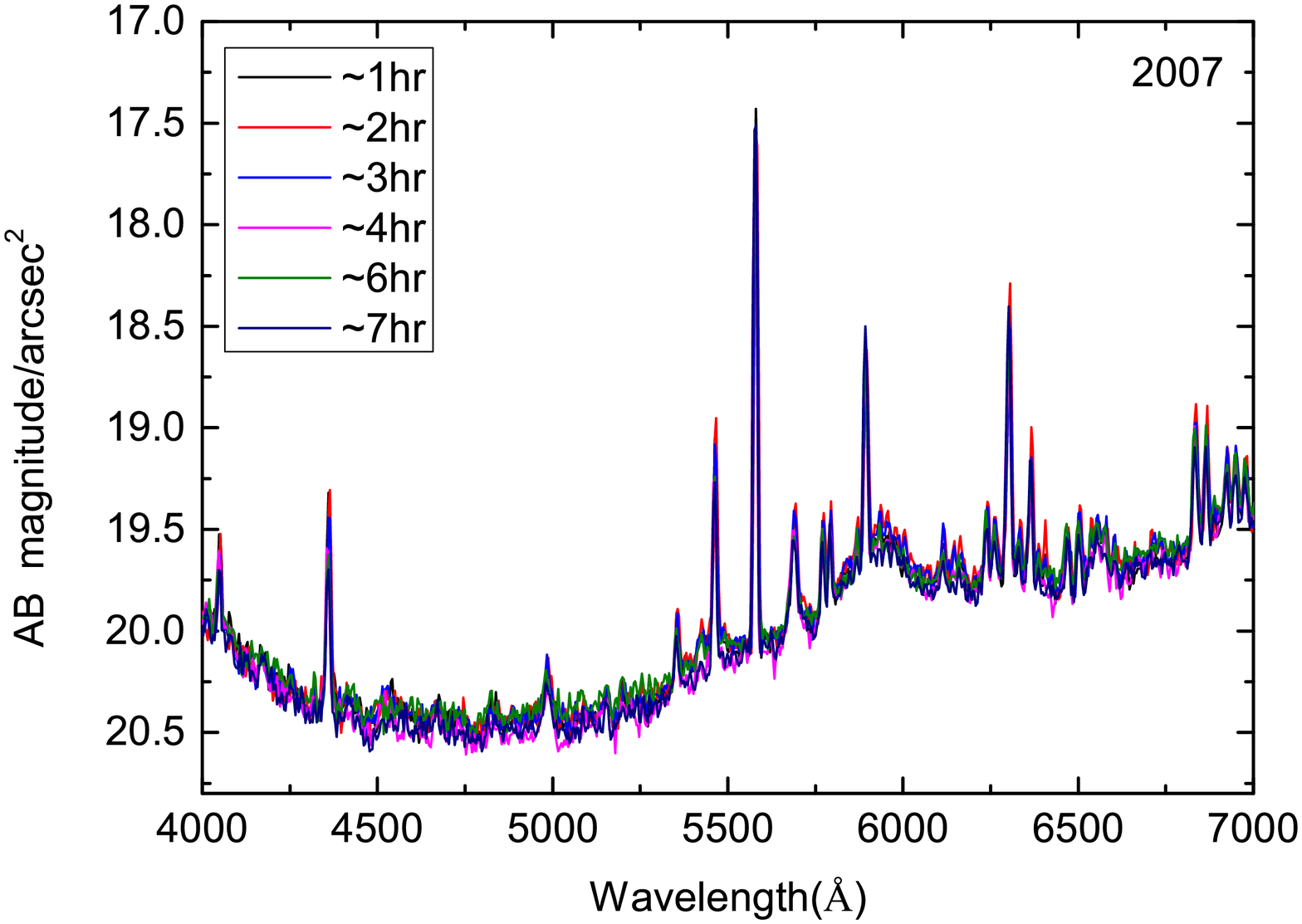}
\includegraphics[angle=0,scale=0.45]{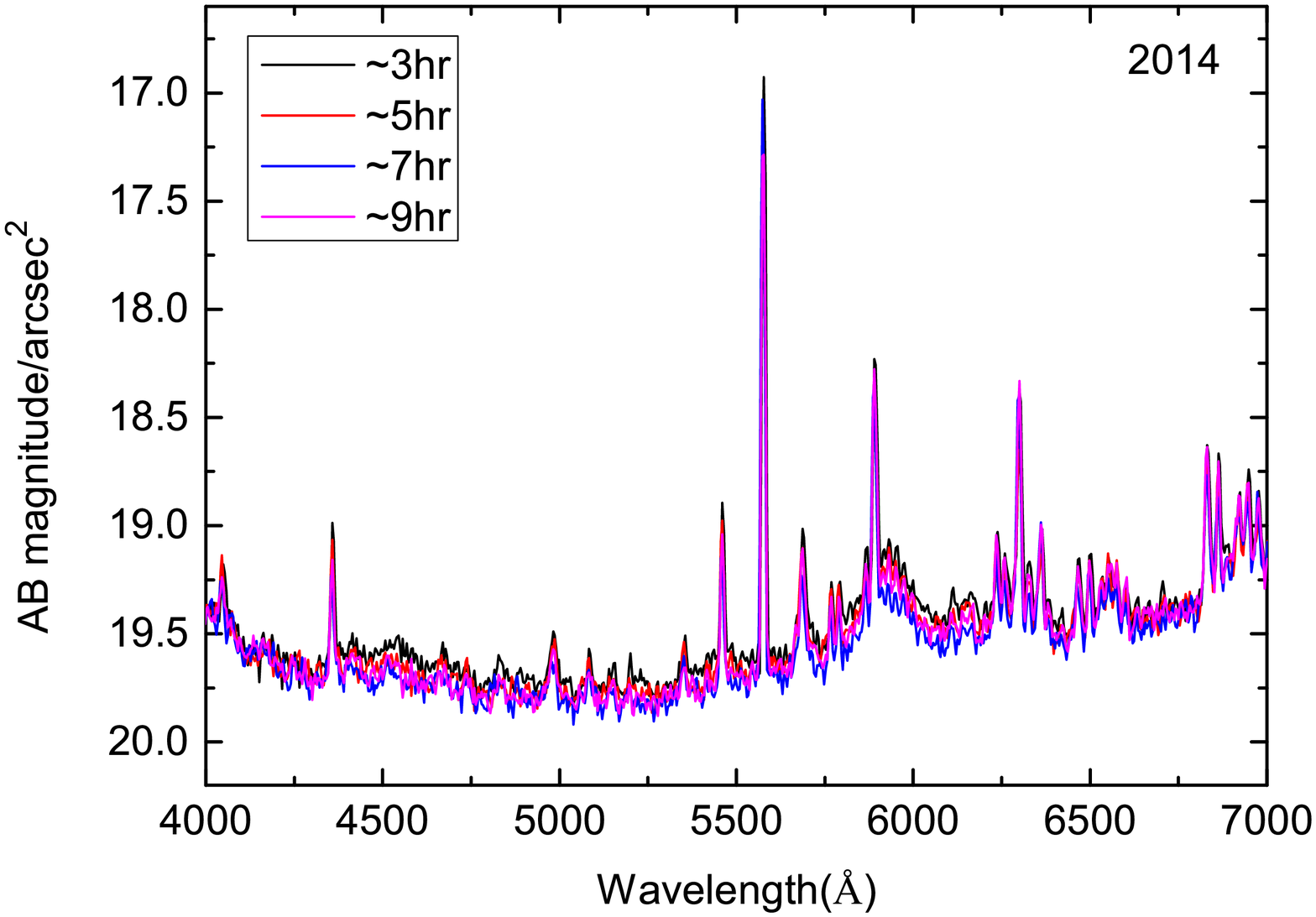}
\hfill \caption{Variation in night sky brightness with time during whole night in 2007 and 2014 are shown. See the electronic edition of the PASP for a color version of this figure.}
\label{fit3}
\end{figure*}

We chose the data one hour after the end of evening astronomical twilight (sun is $18^{\circ}$ below the horizon) when sky becomes completely dark to estimate the night sky brightness and monitored for the whole night in one hour interval to see whether any changes in sky brightness in a single observing dark night. The sky brightness obtained for throughout the night during 2007 is shown in Figure~\ref{fit3}. The altitude of the telescope pointing for all spectra are over $60^{\circ}$. We noticed a little decrement in the sky brightness, but not significant. The same analysis for the data obtained during 2014 also confirms the similar trend.

\begin{figure*}[htbp]
\centering
\includegraphics[angle=0,scale=0.26]{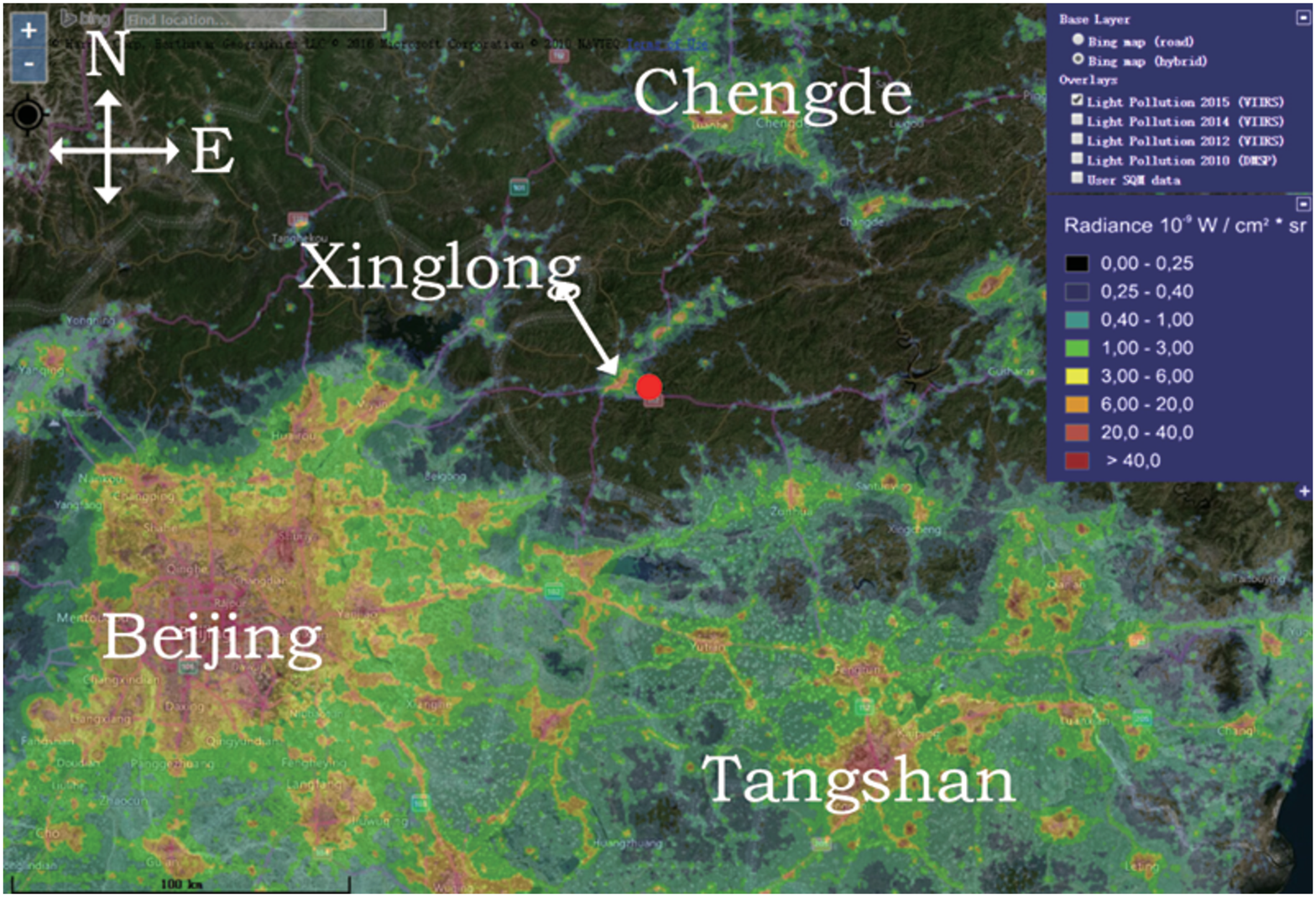}
\includegraphics[angle=0,scale=0.26]{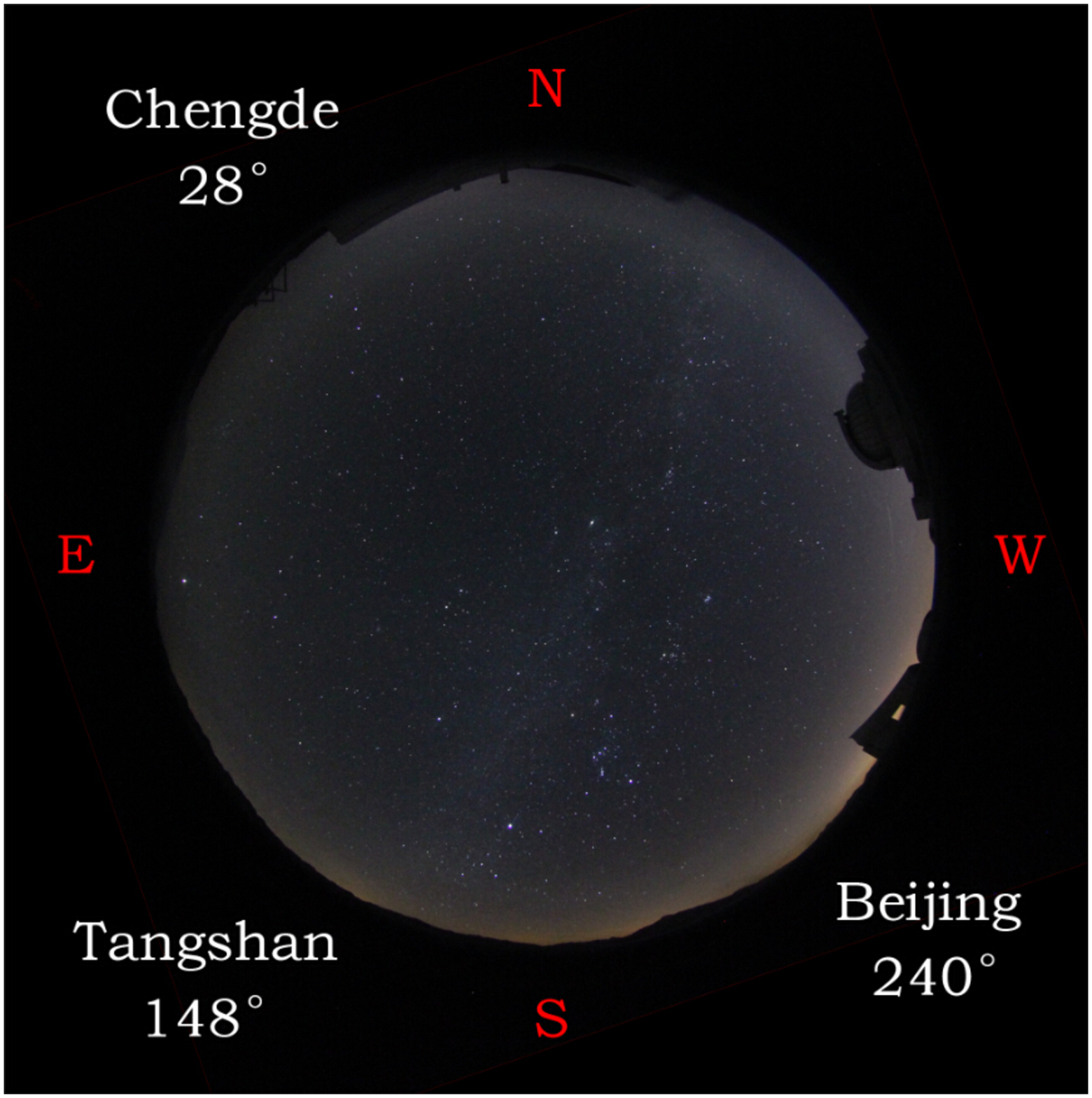}
\hfill \caption{Location of Xinglong observatory with red dot and main surrounding cities and towns are in light pollution map (left panel). All-sky Camera image is mirror reflected in right panel. See the electronic edition of the PASP for a color version of this figure.}
\label{fit4}
\end{figure*}

\begin{figure*}[htbp]
\centering
\includegraphics[angle=0,scale=0.45]{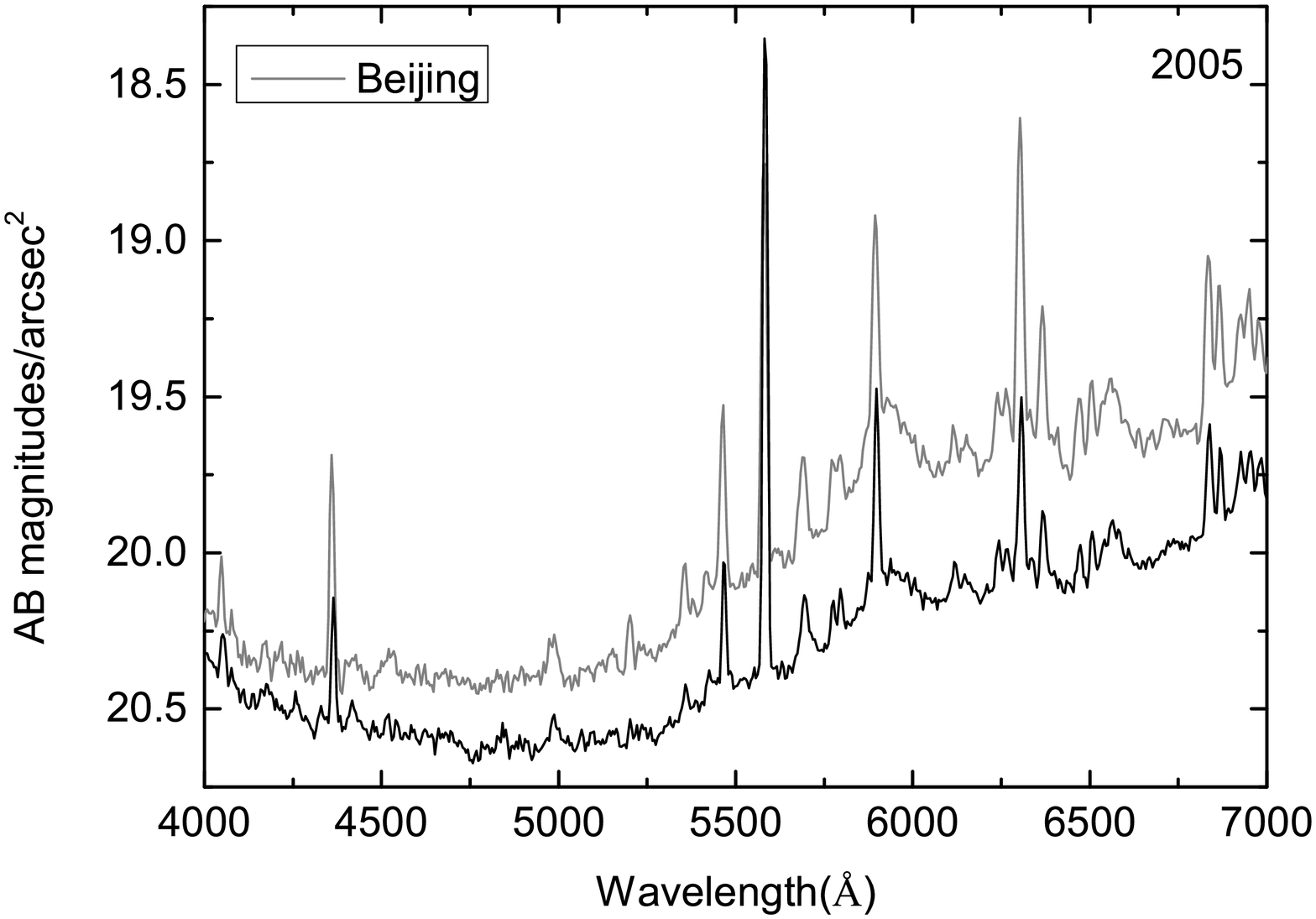}
\includegraphics[angle=0,scale=0.45]{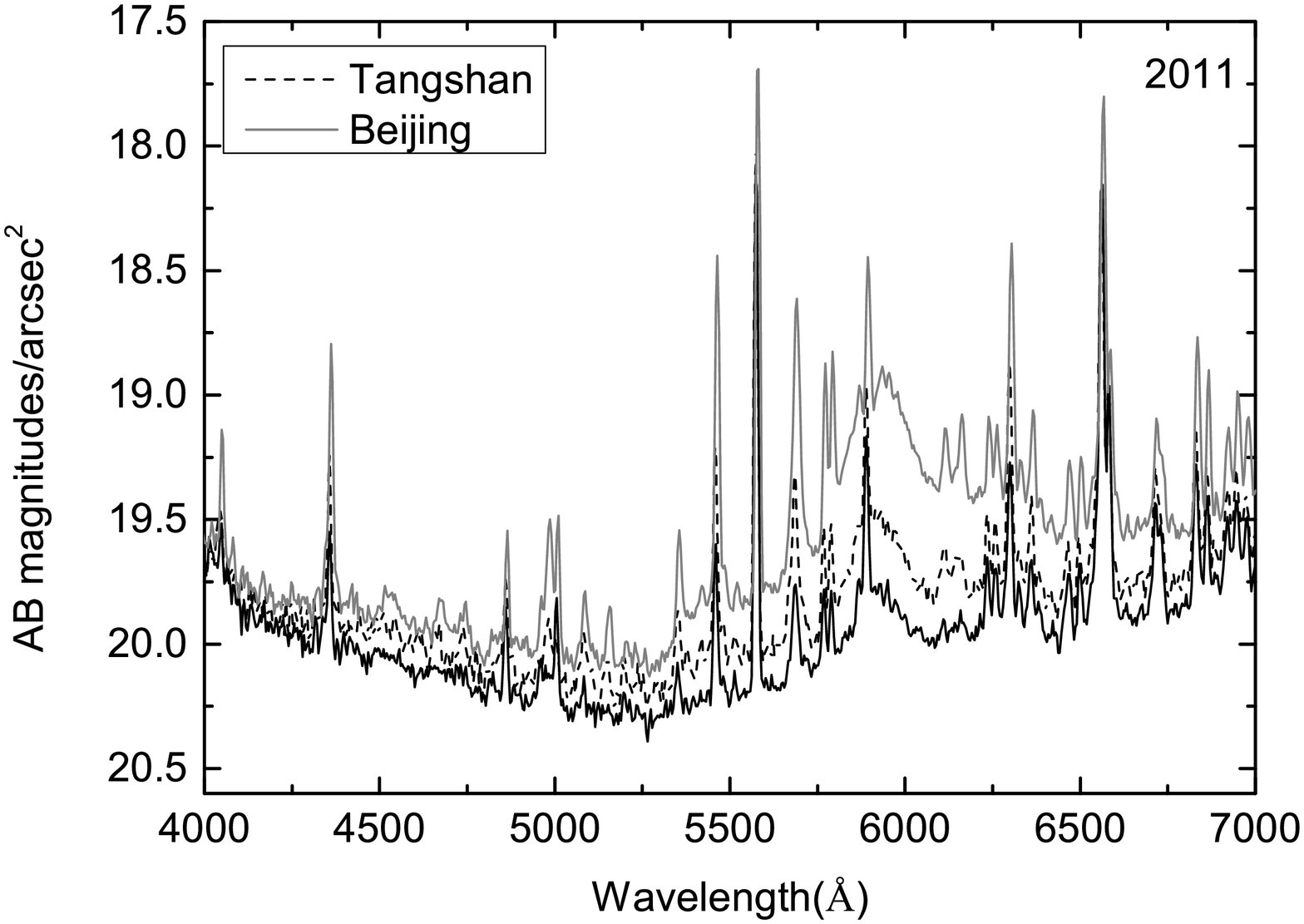}
\hfill \caption{Comparison of night sky spectra at zenith with zenith distance around $50^{\circ}$ towards the direction of Beijing in 2005. A similar comparison of Xinglong Observatory with zenith distance around $60^{\circ}$ in the Beijing and Tangshan direction is shown for 2011.}
\label{fit5}
\end{figure*}

Figure~\ref{fit4} shows the location of cities and towns around Xinglong Observatory in a light pollution map \footnote[3] {The use of data from www.lightpollutionmap.info should be credited to "Jurij Stare, www.lightpollutionmap.info", but because the original data is sourced from Earth Observation Group, NOAA National Geophysical Data Center it should also be credited as such.} and All-sky Camera image of Xinglong Observatory. Note the bright night sky at larger zenith distance is mainly influenced by the surrounding cities. We made quantitative analysis to check the change in night sky brightness at different zenith distance and azimuthal direction and found that night sky is brighter towards South-West where Beijing is located. Relative sky brightness between zenith and a zenith distance around $50^{\circ}$ towards Beijing direction is shown in Figure~\ref{fit5}. For comparison, we have selected data toward Tangshan and showed along with Beijing direction in Figure~\ref{fit5}. This analysis suggests that influence of light pollution from Tangshan is relatively insignificant compare to Beijing. As Xinglong town is located West to Xinglong Observatory and very close compared with other cities, we could find that Xinglong town also has a significant influence of light pollution to Observatory from the image of All-sky Camera.

\begin{figure*}[htbp]
\centering
\includegraphics[angle=0,scale=0.5]{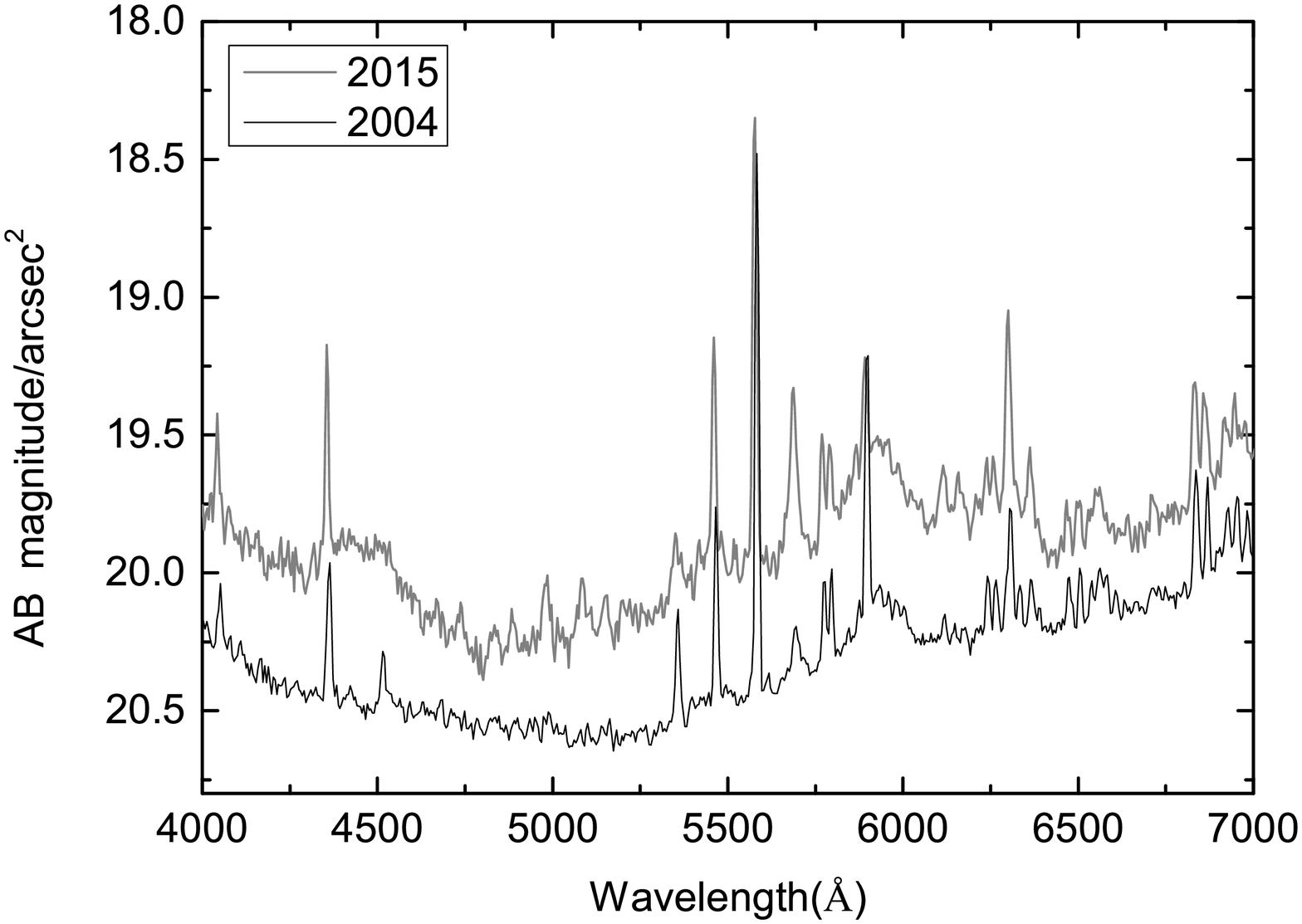}
\hfill \caption{Comparison of night sky spectra of Xinglong Observatory between 2004 and 2015.}
\label{fit6}
\end{figure*}

\begin{figure*}[!htb]
\centering
\includegraphics[angle=0,scale=0.44]{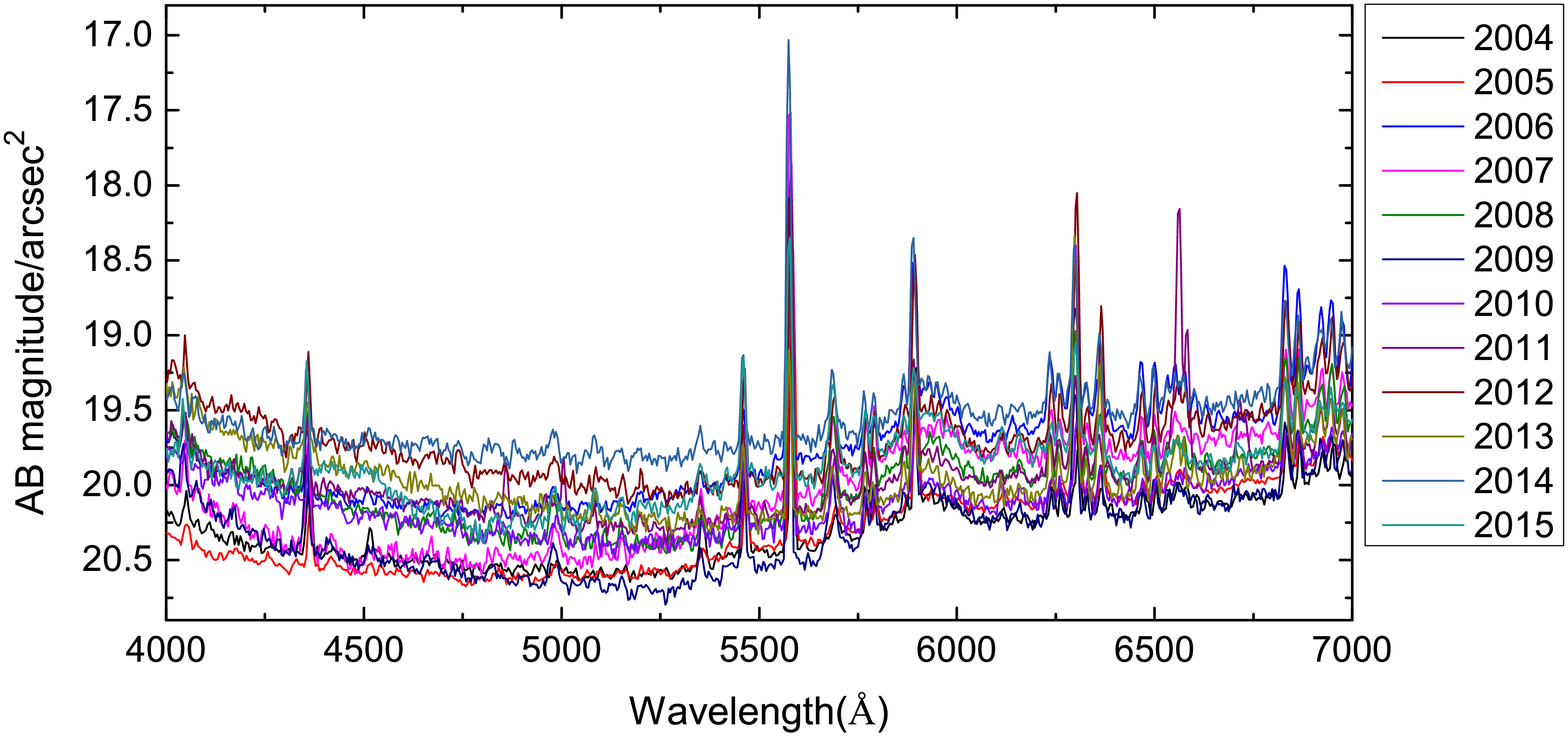}
\hfill \caption{Sample night sky spectra of Xinglong Observatory over twelve years. See the electronic edition of the PASP for a color version of this figure.}
\label{fit7}
\end{figure*}

In order to check the changes in night sky brightness at Xinglong Observatory over past years, representative night sky spectra from 2004 and 2015 are shown in Figure~\ref{fit6} for comparison. We noticed a significant increase in the sky brightness in 2015 compared with 2004. As we concern whether night sky brightness has changed during all these years from 2004 to 2015, we have analyzed the spectra for all the years and showed the sample spectrum from each year in Figure~\ref{fit7}. We could find that the night sky brightness has an increase tendency from 2004 to 2015. Table~1 shows the night sky brightness in broadband B and V during these years, these broadband values were convolved by the corresponding night sky spectrum, we found that the zenith sky brightness over Xinglong Observatory increased slightly between 2004 and 2015. Zenith sky has brightened about 0.5 mag arcsec$^{-2}$ in B and V band.

\begin{table}[htbp]
  \centering
\caption{Statistics of nights sky brightness at Xinglong Observatory from 2004 to 2015.}
    \begin{tabular}{ c c c }
    \hline
    \hline
   &Broadband & (mag arcsec$^{-2}$)\\
    \hline
     Year &  V & B\\
      \hline
     2004 &	20.35 & 20.37\\
     2005 &	20.32 & 20.46\\
     2006 &	19.80 & 19.91\\
     2007 &	19.98 & 20.23\\
     2008 &	20.11 & 19.91\\
     2009 &	20.36 & 20.25\\
     2010 &	20.17 & 20.02\\
     2011 &	20.05 & 19.88\\
     2012 &	19.90 & 19.68\\
     2013 &	20.11 & 19.67\\
     2014 &	19.56 & 19.59\\
     2015 &	19.89 & 19.91\\
    \hline
    \end{tabular}
\end{table}

\section{CONCLUSIONS}

We made an attempt to study the night sky brightness at Xinglong Observatory based on spectroscopic measurements. Overnight monitoring of sky brightness suggests the zenith night sky brightness decrease with time in night, but not significant.
We noticed the strong emission lines from Hg~{\sc i} and Na~{\sc i} in the spectra, apart from natural light emission lines, indicates the influence of light pollution from the usage of Mercury and Sodium lamps in surrounding cities of Xinglong Observatory. Influence of light pollution from the direction of Beijing city is stronger compared with that from the direction of Tangshan city and others. We compared the night sky spectra of twelve years from 2004 to 2015 and found increasing trend in the sky brightness during these years. The convolution of broadband magnitudes suggests that zenith sky has brightened about 0.5 mag arcsec$^{-2}$ in V and B bands. Consecutive spectroscopic measurements in the night sky brightness at Xinglong Observatory in the future is apparently essential, not only for monitoring, but also for scientific reference.

We thank observers and night assistants of 2.16-m telescope for their kind support in obtaining data. This work is partly supported by National Natural Science Foundation of China under grant No.11373003 and National Key Basic Research Program of China (973 Program) No. 2015CB857002. Y.B.K. thanks the Chinese Academy of Sciences Visiting Fellowship for Researchers from Developing Countries for support through grant 2013FFJB0008.

\end{document}